\documentclass[envcountsame]{llncs}

\usepackage[utf8]{inputenc}
\usepackage[T1]{fontenc}
\usepackage{amsmath}
\usepackage{amssymb}
\usepackage{xspace}
\usepackage{boxedminipage}
\usepackage[boxed,ruled]{algorithm}
\usepackage{algorithmic}
\usepackage{paralist}

\renewcommand{\paragraph}[1]{\smallskip\noindent{\bf #1.}\xspace}

\newcommand{\poly}{\operatorname{poly}}

\newcommand{\V}{\mathrm{V}}

\newcommand{\den}{\rho}

\newcommand{\cR}{{\mathcal{R}}}

\newtheorem{observation}[theorem]{Observation}

\def\danupon#1{}
\def\amitabh#1{}
\def\atish#1{}

\title{Dense Subgraphs on Dynamic Networks}

\titlerunning{Dense Subgraphs on Dynamic Networks}

\author{
Atish~{Das~Sarma}\inst{1}
\and Ashwin~Lall\inst{2}
\and Danupon~Nanongkai\inst{3}
\and Amitabh~Trehan\inst{4}\fnmsep\thanks{Supported by a Technion fellowship.}
}

\institute{eBay Research Labs, San Jose, CA, USA. \and Department of Mathematics and Computer Science, Denison University, Granville, OH, USA. \and  University of Vienna, Austria, and Nanyang Technological University, Singapore. \and Information Systems group, Faculty of Industrial Engineering and Management, Technion - Israel Institute of Technology, Haifa, Israel - 32000.}

\authorrunning{A. Das Sarma, A. Lall, D. Nanongkai and A. Trehan}

\tocauthor{A. Das Sarma, A. Lall, D. Nanongkai and A. Trehan}

\begin{document}

\maketitle
\danupon{Don't forget to provide link to full version.}
\begin{abstract}

In  distributed networks, it is often useful for the nodes to be aware of dense subgraphs, e.g., such a dense subgraph could reveal dense subtructures in otherwise sparse graphs (e.g. the World Wide Web or social networks); these might reveal community clusters or dense regions for possibly maintaining good communication infrastructure.
In this work, we address the problem of self-awareness of nodes in a dynamic network with regards to graph density, i.e., we give distributed algorithms for maintaining dense subgraphs that the member nodes are aware of. The only knowledge that the nodes need is that of the \emph{dynamic diameter} $D$, i.e., the maximum number of rounds it takes for a message to traverse the dynamic network. For our work, we consider a model where the number of nodes are fixed, but a powerful adversary can add or remove a limited number of edges from the network at each time step. The communication is by broadcast only and  follows the CONGEST model. Our algorithms are continuously executed on the network, and at any time (after some initialization) each node will be aware if it is part (or not) of a particular dense subgraph. We give algorithms that ($2 + \epsilon$)-approximate the \emph{densest subgraph} and ($3 + \epsilon$)-approximate the \emph{at-least-$k$-densest subgraph} (for a given parameter $k$). Our algorithms work for a wide range of parameter values and run in $O(D\log_{1+\epsilon} n)$ time. Further, a special case of our results also gives the first fully decentralized approximation algorithms for densest and at-least-$k$-densest subgraph problems for static distributed graphs.

\end{abstract}

\section{Introduction}\label{sec:intro}


Density is a very well studied graph property with a wide range of applications stemming from the fact that it is an excellent measure of the strength of inter-connectivity between nodes. While several variants of graph density problems and algorithms have been explored in the classical setting, there is surprisingly little work that addresses this question in the distributed computing framework. This paper focuses on decentralized algorithms for identifying dense subgraphs in dynamic networks.


Finding dense subgraphs has received a great deal of attention in graph algorithms literature because of the robustness of the property. The density of a subgraph only gradually changes when edges come and go in a network, unlike other graph properties such as connectivity that are far more sensitive to perturbation. Density measures the {\em strength} of a set of nodes by the graph induced on them from the overall structure. The power of density lies in locally observing the strength of {\em any} set of nodes, large or small, independent of the entire network.

Dense sugraphs often give key information about the network structure, its evolution and dynamics. To quote~\cite{GibsonKT05}:\emph{``Dense subgraph extraction is therefore a key primitive for any in-depth study of the nature of a large graph''}. Often, dense subgraphs may reveal information about community structure in otherwise sparse graphs e.g. the World Wide Web or social networks. They are good structures for studying the dynamics of a network and have been used, for example, to study link spams~\cite{GibsonKT05}. It is also possible to imagine a scenario where a dynamically evolving peer-to-peer network may want to route traffic through the densest parts of its network to ease congestion; thus, these subgraphs could form the basis of an efficient communication backbone (in combination with other subgraphs selected using appropriate centrality measures).

In this paper, we expand the static CONGEST model \cite{peleg} and consider a dynamic setting where the graph edges may change continually. We present algorithms for approximating the (at least size $k$) densest subgraph in a dynamic graph model to within constant factors. Our algorithms are not only designed to compute size-constrained dense subgraphs, but also track or maintain them through time, thereby allowing the network to be aware of dense subgraphs even as the network changes. They are fully decentralized and adapt well to rapid network failures or modifications. This gives the densest subgraph problem a special status among global graph problems:
while most graph problems are hard to approximate in $o(\sqrt{n})$ time even on static distributed networks of small diameters \cite{DasSarmaHKKNPPW11,NanongkaiSP11,FrischknechtHW12}, the densest subgraph problem can be approximated in polylogarithmic time (in terms of $n$) for small $D$, even in dynamic networks.

We now explain our model for dynamic networks, define density objective s considered in this paper, and state our results.


%



\paragraph{Distributed Computing Model}
Consider an undirected, unweighted, connected $n$-node graph $G = (V, E)$.  Suppose that every node (vertex) hosts a processor with
unbounded computational power (though our algorithms only use time and space polynomial in $n$ at each vertex), but with only local knowledge initially.
We assume that nodes have unique identifiers.
The nodes may accept some additional inputs as specified by the problem at hand. The communication is synchronous, and occurs in discrete pulses, called {\em rounds}. Further, nodes can send messages to each of their neighbors in every round. \amitabh{Check the broadcast assumption too, in the end}
 In our model, all the nodes wake up simultaneously at the beginning of round 1.
 In each round
each node $v$ is allowed to send an arbitrary message subject to the bandwidth constraint of size
$O(\log n)$ bits through any edge $e = (v, u)$ that is adjacent to $v$,
and these messages will arrive at each corresponding neighbor at the end of the current round.
Our model is akin to the standard model of distributed computation known as the
{\em CONGEST model}~\cite{peleg}.
 The message size constraint of CONGEST is very important for large-scale resource-constrained dynamic networks where running time is crucial.

\paragraph{Edge-Dynamic Network Model}
We use the edge deletion/addition model; i.e., we consider a sequence of (undirected) graphs $G_0, G_1, \ldots$ on $n$ nodes, where, for any $t$, $G_t$ denotes the state of the dynamic network $G(V,E)$ at time $t$, where the adversary deletes and/or inserts upto $r$ edges at each step, i.e.,  $E(G_{t+1})= (E(G_{t})\setminus E_U) \cup E_V$, where $E_U \subseteq E(G_{t})$ and $E_V \subseteq E(\overline{G_{t}})$, $|E_U| + |E_V| \leq r$ (where $\overline{G_{t}}$ is the complement graph of $G_{t}$). The edge change rate is denoted by the parameter $r$.

Following the notion in \cite{KuhnOM11}, we define the {\em dynamic diameter} of the dynamic network $G(V,E)$, denoted by $D$, to be the maximum time a message needs to traverse the network at any time.
More formally, dynamic diameter is defined as follows:
\begin{definition}[Dynamic Diameter (Adapted from \cite{KuhnOM11}, Definition 3)]
We say that the dynamic network $G=(V,E)$ has a dynamic diameter of $D$ upto time $t$ if $D$ is the smallest positive integer such that, for all $t' \le t$ and $u,v \in V$, we have $(u,max\{0,t' - D\}) \leadsto (v,t')$, where, for each pair of vertices $x, y$ and times $t_1 \leq t_2$, $(x,t_1) \leadsto (y,t_2)$ means that at time $t_2$ node $y$ can receive direct information, through a chain of messages, originating from node $x$ at time $t_1$.
\end{definition}

Note that the nodes do not need to know the exact dynamic diameter $D$ but only a (loose) approximation to it. For simplicity, we assume henceforth that the nodes know the exact value of $D$.


There are several measures of efficiency of distributed algorithms, but we will concentrate on one of them, specifically, {\em the running time}, that is, the number of rounds of distributed communication. (Note that the computation that is performed by the nodes locally is ``free'', i.e., it does not affect the number of rounds.)




\begin{figure*}[tb]
\caption{The  distributed Edge Insert and Delete Model.}
\label{algo: model-general}
\begin{boxedminipage}{\textwidth}
{\fontsize{10}{10}\selectfont
\begin{algorithmic}
\STATE Each node of $G_0$ is a processor.
\STATE Each processor starts with a list of its neighbors in $G_0$.
\STATE Pre-processing: Processors may exchange messages with their neighbors.
\FOR {$t := 1$ to $T$}
\STATE Adversary deletes and/or inserts upto $r$ edges at each step i.e.  $E(G_{t+1})=(E(G_{t})\setminus E_U) \cup E_V$, where $E_U \subseteq E(G_{t})$ and $E_V \subseteq{E(\overline{G_{t}})}$ (where $\overline{G_{t}}$ is the complement graph of $G_t$).
\IF{edge $(u,v)$ is inserted or edge $(u,v)$ is deleted}
\STATE Nodes $u$ and $v$ may update their information and exchange messages with their neighbors.
\STATE {\bf Computation phase:}
\STATE Nodes may communicate (synchronously, in parallel)
with their immediate neighbors. These messages are never lost or
corrupted, may contain the names of other vertices, and are received by the end of this phase.
\ENDIF
\STATE At the end of this phase, we call the graph $G_t$.
\ENDFOR
\vspace{5pt}
\hrule
\STATE {\bf Success metrics:}
\begin{compactenum}
\item{\bf Approximate Dense Subgraphs:} \emph{Graph $S'_T$:}
 The induced graph of a set $S'_T \subseteq V_T$, s.t., $\den(S'_T) \ge \frac{\den(S^{*}_T)}{\alpha}$, where  $S^{*}_T \subseteq V$, s.t., $\den(S^{*}_T) = \max \den(S_T)$ over all $S_T \subseteq V_T$.
\item{\bf Approximate at-least-k-Dense Subgraphs:} \emph{Graph $S^{k}_T$:} The induced graph of a set  $S^k \subseteq V, |S^k| \geq k$, s.t., $\den(S^k) \ge \frac{\den(S^{k*})}{\alpha}$, where   $S^{k*} \subseteq V, |S^{k*}| \geq k$, s.t., $\den(S^{k*}) = \max \den(S)$ over all $S \subseteq V, |S| \geq k$.

\item{\bf Communication per edge.} The maximum number of bits sent across a single edge in a single recovery round. $O(\log n)$ in CONGEST model.
\item{\bf Computation time.} The maximum total time (rounds) for all nodes to compute their density estimations starting from scratch assuming it takes a message no more than $1$ time unit to traverse any edge and we have unlimited local computational power at each node. 
\end{compactenum}
\end{algorithmic}
} 
\end{boxedminipage}
\end{figure*}


We are interested in algorithms that can compute and maintain an approximate (at-least-$k$) densest subgraph
of the network at all times, after a short initialization time. We say that an algorithm can compute and maintain a solution $P$ in time $T$ if it can compute the solution in $T$ rounds and can maintain a solution at all times after time $T$, even as the network changes dynamically.

\subsection{Problem definition}


Let $G =(V,E)$ be an undirected graph and $S \subseteq \V$ be a set of nodes. Let us define the following:

\paragraph{Graph Density}
The density of a graph $G(V, E)$ is defined as $|E|/|V|$.

\paragraph{SubGraph Density}
The density of a subgraph defined by a subset of nodes $S$ of $V(G)$ is defined as the density of the induced subgraph. We will use $\den(S)$ to denote the density of the subgraph induced by $S$. Therefore, $\den(S) = \frac{|E(S)|}{|S|}$. Here $E(S)$ is the subset of edges $(u, v)$ of $E$ where $u\in S$ and $v\in S$.  In particular, when talking about the density of a subgraph defined by a set of vertices $S$ induced on $G$, we use the notation $\den_G(S)$. We also use $\den_t(S)$ to denote $\den_{G_t}(S)$. When clear from context, we omit the subscript $G$.

The problem we address in this paper is to construct distributed algorithms to discover the following:
\begin{compactitem}
\item{\bf (Approximate) Densest subgraphs:} The densest subgraph problem is to find a set $S^{*} \subseteq V$, s.t. $\den(S^{*}) = \max \den(S)$ over all $S \subseteq V$. A $\alpha$-approximate solution $S'$ will be a set $S' \subseteq V$, s.t. $\den(S') \ge \frac{\den(S^{*})}{\alpha}$.
\item {\bf (Approximate) At-least-$k$-densest subgraphs:} The densest at-least-$k$-subgraph problem is the previous problem restricted to sets of size at least $k$, i.e., to find a set $S^{k*} \subseteq V, |S^{k*}| \geq k$, s.t.\ $\den(S^{k*}) = \max \den(S)$ over all $S \subseteq V, |S| \geq k$. A $\alpha$-approximate solution $S^k$ will be a set $S^k \subseteq V, |S^k| \geq k$, s.t. $\den(S^k) \ge \frac{\den(S^{k*})}{\alpha}$.
\end{compactitem}

In the distributed setting, we require that every node knows whether it is in the solution $S'$ or $S^k$ or not. We note that the latter problem is {\sf NP}-Complete, and thus it is crucial to consider approximation algorithms. The former problem can be solved {\em exactly} in polynomial time in the centralized setting, and it is an interesting open problem whether there is an exact distributed algorithm that runs in $O(D\poly\log n)$ time, even in static networks.

\subsection{Our Results}

We give approximation algorithms for the densest and at-least-$k$-densest subgraph problems which are efficient even on dynamic distributed networks. In particular, we develop an algorithm that, for a fixed constant $c$ and any $\epsilon > 0$, $(2+\epsilon)$-approximates the densest subgraph in $O(D \log_{1+\epsilon} n)$ time provided that the densest subgraph has high density, i.e., it has a density at least $(cDr\log n)/\epsilon$ (recall that $r$ and $D$ are the change rate and dynamic diameter of dynamic networks, respectively). We also develop a $(3+\epsilon)$-approximation algorithm for the at-least-$k$-densest subgraph problem with the same running time, provided that the value of the density of the at-least-$k$-densest subgraph is at least $(cDr\log n)/k\epsilon$. We state these theorems in a simplified form and some corollaries below.
Below, $\epsilon$ can be set as any arbitrarily small constant. We note again that at the end of our algorithms, every node knows whether they are in the returned subgraph or not.

\begin{theorem}
There exists a distributed algorithm that for any dynamic graph with dynamic diameter $D$ and parameter $r$ returns a subgraph at time $t$ such that, w.h.p., the density of the returned subgraph is a $(2+\epsilon)$-approximation to the density of the densest subgraph at time $t$ if the densest subgraph has density at least $\Omega(Dr\log n)$.
\end{theorem}

\begin{theorem}
There exists a distributed algorithm that for any dynamic graph with dynamic diameter $D$ and parameter $r$ returns a subgraph of size at least $k$ at time $t$ such that, w.h.p., the density of the returned subgraph is a $(3+\epsilon)$-approximation to the density of the densest at least $k$ subgraph at time $t$ if the densest at least $k$ subgraph has density at least $\Omega(Dr\log n/k)$.
\end{theorem}

We mention two special cases of these theorems informally below. We prove the most general theorem statements depending on the parameters $r$ and $D$ in Section~\ref{sec:approx}.

\begin{corollary}
Given a dynamic graph with dynamic diameter $O(\log n)$ and a rate of change $r = O(\log^{\alpha} n)$ for some constant $\alpha$ (i.e. $r$ is poly-logarithmic in $n$), there is a distributed algorithm that at any time $t$ can return, w.h.p., a $(2+\epsilon)$-approximation of densest subgraph at time $t$ if the densest subgraph has density at time $t$ at least $\Omega(\log^{\alpha +2} n)$.
\end{corollary}

\begin{corollary}
Given a dynamic graph with dynamic diameter $O(\log n)$ and a rate of change $r = O(\log^{\alpha} n)$ for some constant $\alpha$ (i.e. $r$ is poly-logarithmic in $n$), there is a distributed algorithm that at any time $t$ can return, w.h.p., a $(3+\epsilon)$-approximation of $k$-densest subgraph at time $t$ if the $k$-densest subgraph has density at time $t$ at least $\Omega(\log^{\alpha +2} n/k)$.
\end{corollary}

Our algorithms follow the main ideas of centralized approximation algorithms \cite{KS,AC,Charikar00}\danupon{We didn't mention Charikar at all in the related work!}. These centralized algorithms cannot be efficiently implemented even on static distributed networks. We show how some ideas of these algorithms can be turned into time-efficient distributed algorithms with a small increase in the approximation guarantees. Similar ideas have been independently discovered and used to obtain efficient streaming and MapReduce algorithms by Bahmani et al. \cite{BahmaniKV12}.

Notice that this is already a wide range of parameter values for which our results are interesting, since the density of densest subgraphs can be as large as $\Omega(n)$ while the diameter in peer-to-peer networks is typically $O(\log n)$, and the parameter $r$ depends on the stability of the network. A caveat, though, is that in the theorems above, $D$ refers to the flooding time of the dynamic network, and not the diameter of any specific snapshot - understanding a relationship between these quantities remains open.

Further, our general theorems also imply the following for static graphs (by simply setting $r = 0$). No such results were known in the distributed setting even for static graphs.

\begin{corollary}
In a static graph, there is a distributed algorithm that obtains, w.h.p., $(2+\epsilon)$-approximation to the densest subgraph problem in $O(D\log n)$ rounds of the CONGEST model.
\end{corollary}

\begin{corollary}
In a static  graph, there is a distributed  algorithm that obtains, w.h.p, $(3+\epsilon)$-approximation to the $k$-densest subgraph problem in $O(D\log n)$ rounds of the CONGEST model.

\end{corollary}

Notice that this is an unconditional guarantee for static graphs (i.e. does not require any bound on the density of the optimal) and is the first distributed algorithm for these problems in the CONGEST model.

Back to dynamic graphs, in addition to computing the $(2+\epsilon)$-approximated densest and $(3+\epsilon)$-approximated at-least-$k$-densest subgraphs, our algorithm can also {\em maintain} them {\em at all times} with high probability. This means that, at all times (except for a short initialization period), all nodes are aware of whether they are part of the approximated at-least-$k$ densest subgraphs, for all $k$.

Even though we assume that all the nodes know the value $D$, all our algorithms work if some upper-bound $D'$ of $D$ is known instead; all the algorithms and analysis work identically using $D'$ rather than $D$.

\paragraph{Organization} Our algorithms are described in Section~\ref{sec:algo} and the approximation guarantees are proved in Section~\ref{sec:approx}. We mention related work at the end of the paper in Section~\ref{sec:relatedwork}.
%
%
%
%

\section{Algorithm}\label{sec:algo}

\subsection{Main Algorithm}
\label{sec:main}

The nature of our algorithm is such that we {\em continuously} maintain an approximation to the densest subgraph in the dynamic network. At any time, after a short initialization period, any node knows whether it is a member of the output subgraph of our algorithm. In this section, we give the description of the algorithm and fully specify the behavior of each of the nodes in the network. The running time analysis and the approximation guarantees are deferred to the following sections.

Our main protocol for maintaining a dense subgraph is given in Algorithm~\ref{algo:maintain}. It maintains a family of  $p=O(\log_{1+\epsilon} n)$ candidates for the densest subgraph $\mathcal{F}=\{V_0, V_1, \ldots, V_p\}$, where $V_0=V(G)$, $V_i\subseteq V_{i-1}$ for all $i$, along with an approximation of the number of nodes and edges in each graph $\cR=\{(m_0, n_0), \ldots, (m_p, n_p)\}$, where each $m_i$ and $n_i$ are the approximate number of edges and nodes, respectively, of the subgraph of $G_t$ (the current graph) induced by $V_i$. The algorithm works in phases in which it estimates the size of the current subgraph $V_j$ and the number of edges in it using the algorithms discussed in the following subsection. At the end of the phase it computes the next subgraph $V_{j+1}$ using a criterion in Line 9 of Algorithm~\ref{algo:maintain} (explained further in Section~\ref{sec:approx}). After $p$ such rounds, the algorithm has all the information it needs to output an approximation to the densest subgraph. This process is repeated continuously, and the solution is computed from the last complete family of graphs (i.e., complete computation of $p$ subgraphs).

\begin{algorithm*}[tb]
\caption{{\sc Maintain}($\epsilon$)}
\label{algo:maintain}

{\bf Input:} $1\geq \epsilon>0$

{\bf Output:} The algorithm maintains a family of sets of nodes $\mathcal{F}=\{V_0, V_1, \ldots, V_p\}$ and induced graph sizes $\cR=\{(m_0, n_0), (m_1, n_1), \ldots, (m_p, n_p)\}$.

\begin{algorithmic}[1]

\STATE Let $\delta=\epsilon/24$.

\STATE Let $j = 0$. Let $V_0=V$ (i.e., we mark every node as in $V_0$).

\REPEAT

\STATE Compute $n_j$, a $(1+\delta)$-approximation of $|V_j|$ (i.e., $(1+\delta)|V_j|\geq n_j\geq (1-\delta)|V_j|$). At the end of this step every node knows $n_j$. See Algorithms~\ref{algo:kuhn} and~\ref{algo:count nodes} for detailed implementation.

\IF{$n_j=0$}

\STATE Let $j = 0$. 
(Note that we do not recompute $n_0$.)

\ENDIF

\STATE Let $G_t$ be the network at the beginning of this step. Let $H_t$ be the subgraph of $G_t$ induced by $V_{j}$. We compute $m_{j}$, the $(1+\delta)$-approximation of the number of edges in $H_t$ (i.e., $(1+\delta)|E(H_t)|\geq m_{j}\geq (1-\delta)|E(H_t)|$). At the end of this step every node knows $m_{j}$. See Algorithm~\ref{algo:count edges} for detailed implementation.


\STATE Let $G_{t'}$ be the network at the beginning of this step.  Let $H_{t'}$ be the subgraph of $G_{t'}$ induced by $V_{j}$. Let $V_{j+1}$ be the set of nodes in $V_{j}$ whose degree in $H_{t'}$ is at least $(1+\delta)m_j/n_j$. At the end of this step, every node knows whether it is in $V_{j+1}$ or not.


\STATE Let $j=j+1$.

\UNTIL{forever}

\end{algorithmic}
\end{algorithm*}

At any time, the densest subgraph can be computed using the steps outlined in Algorithm~\ref{algo:densest}. This procedure works simply by picking the subgraph with the highest density, even if the size of this subgraph is less than $k$. If the graph turns out to be less than size $k$, we pad it by having the rest of the nodes run a distributed procedure to elect appropriately many nodes to add to the subgraph and get its size up to at least $k$.

Any time a densest subgraph query is initiated in the network, the nodes simply run Algorithm~\ref{algo:densest} based on the subgraphs continuously being maintained by Algorithm~\ref{algo:maintain}, and compute which of them are in the approximation solution. At the end of this query, each node is aware of whether it is in the approximate densest subgraph or not.

\begin{algorithm*}[htbp]
\caption{{\sc Densest Subgraph}($k$)}
\label{algo:densest}

{\bf Input:} $k$, the parameter for the densest at-least-$k$ subgraph problem, the algorithm {\sc Maintain($\epsilon$)} (cf. Algorithm~\ref{algo:maintain}), and its parameter notations.


{\bf Output:} The algorithm outputs a set of nodes $V_i\cup\hat{V}$ (every node knows whether it is in the set or not) such that $|V_i\cup\hat{V}|\geq k$.


\begin{algorithmic}[1]
\STATE Let $i=\max_{i} m_i/\max(k, n_i)\,.$
\IF{$n_i<(1+\delta)k$}
\STATE Let $\Delta=(1+\delta)k-n_i$. (Every node can compute $\Delta$ locally.)
\REPEAT
\STATE Every node not in $V_i$ locally flips a coin which is head with probability $\Delta/n_0$.
\STATE Let $\hat{V}$ be the set of nodes whose coins return heads.
\STATE Approximately count the number of nodes in $\hat{V}$ using the algorithm {\sc Approx-Size-Estimation} discussed in Section~\ref{sec:counting} with error parameter $\delta$ passed to {\sc Count Edges} under it. Let $\Delta'$ be the result returned. (Note that $\Delta'/(1+\delta)\leq |\hat{V}|\leq (1+\delta)\Delta'$ w.h.p.)
\UNTIL{$(1+\delta)\Delta\leq \Delta'\leq (1+2\delta)\Delta$}
\ENDIF
\RETURN $V_i\cup \hat{V}$
\end{algorithmic}
\end{algorithm*}


\subsection{Approximating the number of nodes and edges}\label{sec:counting}

Our algorithms make use of an operation in which the number of nodes and edges in a given subgraph need to be computed. We just mention the algorithm idea here and present the detailed algorithm in Appendix~\ref{sec:count}.

\paragraph{Algorithm {\sc Approx-Size-Estimation}} We achieve this in $O(D)$ rounds using a modified version of an algorithm from~\cite{KuhnLO10}. Their algorithm allows for approximate counting of the size of a dynamic network with high probability. We modify it to work for any subgraph that we are interested in. We also show how it can be used to approximate the number of edges in this subgraph at a given time. In the interest of space, these results can be found in Appendix~\ref{sec:count} described under algorithms {\sc RandomixedApproximateCounting}, {\sc Count Nodes}, and {\sc Count Edges}.


%
%
%
%
%
%

\section{Analysis}
\label{sec:approx}

We analyze approximation ratios of the algorithm presented in Section~\ref{sec:algo}, the guarantee depending on parameters of the algorithm. We divide the analysis into two parts: the first part is for the densest subgraph problem and the second for the at-least-$k$ densest subgraph problem. Although the second part subsumes the first part (if we ignore the value of constant approximation ratio), we present the first part since it has a simpler idea and a better approximation ratio.

\subsection{Analysis for the densest subgraph problem}

\begin{theorem}\label{theorem:approx densest}
Let $t$ be the time Algorithm~\ref{algo:densest} finishes, $V_i$ be the output of the algorithm, $H^*$ be the optimal solution and $T$ be the time of one round of Algorithm~\ref{algo:maintain} and \ref{algo:densest} (i.e., $T=cD\log_{1+\epsilon} n$ for some constant $c$). If $\rho_t(H^*)\geq 24Tr/\epsilon$ then Algorithm~\ref{algo:densest} gives, w.h.p., a $(2+\epsilon)$-approximation, i.e.,
$$\rho_t(V_i)\geq \rho_t(H^*)/(2+\epsilon)\,.$$
\end{theorem}


The rest of this subsection is devoted to proving the above theorem. 
Let $t$, $V_i$ and $H^*$ be as in the theorem statement (note that $\hat{V}$ in Algorithm~\ref{algo:densest} is empty when $k=0$). Let $t'$ be the time that $V_i$ is last computed by Algorithm~\ref{algo:maintain}. Let $t''$ be the time Algorithm~\ref{algo:maintain} starts counting the number of edges in $V_i$.
We prove the theorem using the following lemmas. The main idea is to first lower bound $\rho_{t''}(V_i)$ using $\rho_{t'}(H^*)$ and then use it to obtain a lower bound for $\rho_{t'}(V_i)$ in terms of $\rho_t(H^*)$. Finally, the proof is completed by lower bounding $\rho_t(V_i)$ in terms of $\rho_{t'}(V_i)$.


\begin{lemma}\label{lem: t prime prime to OPT}
$\rho_{t''}(V_i)>\frac{1-\delta}{2(1+\delta)^2}\rho_{t'}(H^*).$
\end{lemma}
\begin{proof}
Let $H'$ be the densest subgraph of $G_{t'}$. Note that
\begin{align}\label{eq:densest one}
\rho_{t'}(H^*)\leq \rho_{t'}(H')\,.
\end{align}
%
%
Let $i^*$ be the smallest index such that $V(H')\subseteq V_{i^*}$ and  $V(H')\not\subseteq V_{i^*+1}$.  Note that $i^*$ exists since the algorithm repeats until we get $V_j=\emptyset$. Let $v$ be any vertex in $V(H')\setminus V_{i^*}$. Let $H_{t', i}$ be the subgraph of $G_{t'}$ induced by nodes in $V_i$. Note that
\begin{align}\label{eq:densest two}
\rho_{t'}(H')\leq 2 \deg_{H'}(v) \leq 2\deg_{H_{t', i}}(v)\,. 
\end{align}
%
The first inequality is because we can otherwise remove $v$ from $H'$ and get a subgraph of $G_{t'}$ that has a higher density than $H'$. The second inequality is because $H'\subseteq H_{t', i}$.
Since $v$ is removed from $V_{i^*}$,
\begin{align}\label{eq:densest three}
\deg_{H_{t',i}}(v)< (1+\delta) \frac{m_{i^*}}{n_{i^*}},
\end{align}
where $\delta=\epsilon/24$ as in Algorithm~\ref{algo:maintain}.
By the definition of $V_i$,
\begin{align}\label{eq:densest four}
\frac{m_{i^*}}{n_{i^*}}\leq \frac{m_i}{n_i}\,.
\end{align}
Note that $t-t''\leq T$ by the definition of $T$. Note also that $n_{i}\geq (1-\delta) |V_{i}|$ and $m_{i}\leq (1+\delta) |E_{t''}(V_{i})|$ with high probability\danupon{Need proof from previous section.}. It follows that
\begin{align}\label{eq:densest five}
\frac{m_{i}}{n_{i}}\leq \frac{1+\delta}{1-\delta}\rho_{t''}(V_{i})\,.
\end{align}
%
%
Combining Eq.\eqref{eq:densest one}-\eqref{eq:densest five}, we get $\rho_{t'}(H^*)<2\frac{(1+\delta)^2}{1-\delta}\rho_{t''}(V_i)$ and thus the lemma.
\end{proof}

We now make the following observation:

\begin{observation}
\label{obs:43}
$\rho_{t'}(H^*)\geq (1-\delta)\rho_t(H^*)\,.$
\end{observation}
\begin{proof}
Note that $t-t'\leq T$ and thus $E_t(H^*)-E_{t'}(H^*)\leq Tr$. Since $\rho_t(H^*)\geq Tr/\delta$,
$\rho_{t'}(H^*)  \geq \frac{\rho_{t}(H^*)\cdot |V(H^*)|-T r}{|V(H^*)|}
 \geq \rho_{t}(H^*)-Tr
 > (1-\delta)\rho_{t}(H^*)\,.$
\end{proof}

We now combine the above Lemma~\ref{lem: t prime prime to OPT} and Observation~\ref{obs:43} to obtain the following lemma:


\begin{lemma}
\label{lem:45}
$\rho_{t'}(V_i) > (\frac{(1-\delta)^2}{2(1+\delta)^2}-\delta)\rho_{t}(H^*)\,.$
\end{lemma}
\begin{proof}
By directly combining Lemma~\ref{lem: t prime prime to OPT} and Observation~\ref{obs:43} we get the following:
$$\rho_{t''}(V_{i})> \frac{(1-\delta)^2}{2(1+\delta)^2}\rho_{t}(H^*)\geq \frac{(1-\delta)^2}{2(1+\delta)^2\delta} Tr\,.$$
Moreover, observe that there are at most $Tr$ edges removed from $V_i$ in total, i.e., $E_{t''}(V_i)-E_{t}(\
V_i)\leq Tr$. Thus
\begin{align*}
\rho_{t'}(V_i) &\geq \frac{\rho_{t''}(V_i)\cdot |V_i|-T r}{|V_i|}
 \geq \rho_{t''}(V_i)-Tr
 > \left(1-\frac{2(1+\delta)^2\delta}{(1-\delta)^2}\right)\rho_{t''}(V_i)\\
& > \left(1-\frac{2(1+\delta)^2\delta}{(1-\delta)^2}\right)\left(\frac{(1-\delta)^2}{2(1+\delta)^2}\rho_{t}(H^*)\right) 
 = \left(\frac{(1-\delta)^2}{2(1+\delta)^2}-\delta\right)\rho_{t}(H^*)\,.
\end{align*}
%
\end{proof}


We are now ready to prove the theorem. 

\begin{proof}[Proof of Theorem~\ref{theorem:approx densest}]
Note that $t-t'\leq T$ and thus $E_{t'}(V_i)-E_{t}(V_i)\leq Tr$. Note that
$\rho_{t'}(V_i)> \beta\rho_{t}(H^*)\geq \beta Tr/\delta,$
where $\beta=\frac{(1-\delta)^2}{2(1+\delta)^2}-\delta$. We have
\begin{align*}
\rho_{t}(V_i)  \geq \frac{\rho_{t'}(V_i)\cdot |V_i|-T r}{|V_i|}
 \geq \rho_{t'}(V_i)-Tr
 > (1-\frac{\delta}{\beta})\rho_{t'}(V_i).
\end{align*}

Now using Lemma~\ref{lem:45} and the value of $\beta$, we get the following:
\begin{align*}
\rho_{t}(V_i) > (1-\frac{\delta}{\beta})\beta\rho_{t}(H^*) 
 = (\beta-\delta)\rho_{t}(H^*)
 = \left(\frac{(1-\delta)^2}{2(1+\delta)^2}-2\delta\right)\rho_t(H^*).
\end{align*}
The theorem follows by observing that  $\frac{(1-\delta)^2}{2(1+\delta)^2}-2\delta\geq \frac{1}{2+\epsilon}$ for any $\epsilon\leq 1$ and $\delta \geq \epsilon/24$.
\end{proof}

\subsection{Analysis for the at-least-$k$ densest subgraph problem}

\begin{theorem}\label{thm:atleastktheorem}
Let $t$ be the time Algorithm~\ref{algo:densest} finishes, $V_i \cup \hat{V}$ be the output of the algorithm, $H^*$ be the optimal solution and $T$ be the time of one iteration of Algorithm~\ref{algo:maintain} and Algorithm~\ref{algo:densest} (so $T=O(D\log_{1+\epsilon} n)$). If $k\rho_t(H^*)\geq 24Tr/\epsilon$ then Algorithm~\ref{algo:densest} returns a set $V_i\cup \hat{V}$ of size at least $k$ that is, w.h.p., a $(3+\epsilon)$-approximated solution, i.e.,
$$\rho_t(V_i\cup \hat{V})\geq \rho_t(H^*)/(3+\epsilon)\,.$$
\end{theorem}

The proof of this theorem is placed in Appendix~\ref{sec:app main two}, and we just mention the main idea here. The proof follows a similar framework as that of Theorem~\ref{theorem:approx densest}.

Let $t$, $V_i$ and $H^*$ be as in the theorem statement. Let $t'$ be the time that $V_i$ is last computed by Algorithm~\ref{algo:maintain}. Let $t''$ be the time Algorithm~\ref{algo:maintain} starts counting the number of edges in $V_i$. The crucial difference here is to obtain a strong lower bound for $\rho_{t''}(V_i\cup \hat{V})$ in terms of $\rho_{t'}(H^*)$ and $\rho_t(H^*)$. This is then translated to a lower bound on $\rho_{t'}(V_i\cup \hat{V})$ and subsequently $\rho_{t}(V_i\cup \hat{V})$ to complete the proof. The crucial lemma and its proof turn out to be more involved than that of the densest subgraph theorem and the case-based analysis is detailed in Appendix~\ref{sec:app main two}.


\subsection{Running Time Analysis}\label{sec:time}

In this section we analyze the time that it takes for the nodes to generate an approximation to the densest subgraph. Algorithm~\ref{algo:maintain} continuously runs this procedure so that it always maintains an approximation that is guaranteed to be near-optimal since we assume that the network does not change too quickly. 
The time that it takes for Algorithm~\ref{algo:maintain} to compute a complete family of subgraphs is simply $O(Dp) = O(D\log_{1+\epsilon}{n})$ since there are $p = O(\log_{1+\epsilon} n)$ rounds (Section~\ref{sec:main}), each of which is completed in $O(D)$ time (Section~\ref{sec:counting}). Note that step 9 of Algorithm~\ref{algo:maintain} can be done in a single round since every node already knows $m_{j}/n_{j}$ and can easily check, in one round, the number of neighbors in $G_{t'}$ that are in $V_{j}$.

When the nodes need to compute an approximation to the at-least-$k$-densest subgraph in Algorithm~\ref{algo:densest}, they can do so by choosing the densest subgraph among the last complete family of subgraphs found by Algorithm~\ref{algo:maintain}. Unfortunately, there is no guarantee that the densest such graph has at least $k$ nodes in it, so we fix this via padding. The subgraph is padded to contain at least $k$ nodes by having each node that is not part of the subgraph attempt to join the subgraph with an appropriate probability. It can be shown via Chernoff bounds that, with high probability, within $O(\log{n})$ such attempts there are enough nodes added to the subgraph to get its size to at least $k$. As a result, Algorithm~\ref{algo:densest} runs in $O(D\log{n})$ time. 

\section{Related Work}\label{sec:relatedwork}

The problem of finding size-bounded densest subgraphs has been studied extensively in the classical setting. Finding a maximum density subgraph in an undirected graph can be solved in polynomial time~\cite{G84,L}. However, the problem becomes NP-hard when a size restriction is enforced. In particular, finding a maximum density subgraph of size exactly $k$ is NP-hard~\cite{AHI,FKP} and no approximation scheme exists under a reasonable complexity assumption~\cite{K}. Recently Bhaskara et al.~\cite{BCVGZ12} showed integrality gaps for SDP relaxations of this problem.
Khuller and Saha~\cite{KS} considered the problem of finding densest subgraphs with size restrictions and showed that these are NP-hard. Khuller and Saha~\cite{KS} and also Andersen and Chellapilla~\cite{AC} gave constant factor approximation algorithms. Some of our algorithms are based on of those presented in~\cite{KS}.

Our work differs from the above mentioned ones in that we address the issues in a dynamic setting, i.e., where edges of the network change over time.
Dynamic network topology and fault tolerance have always been core concerns of distributed computing~\cite{Attiya-WelchBook,lynch}. There are many models and a large volume of work in this area.
A notable recent model is the dynamic graph model introduced by Kuhn, Lynch and Oshman in \cite{KuhnLO10}. They introduced a  stability property called $T$-interval connectivity (for $T\ge 1$) which stipulates the existence of a stable connected spanning subgraph for every $T$ rounds. Though our models are not fully comparable (we allow our networks to get temporarily disconnected as long as messages eventually make their way through it),
the graphs generated by our model are similar to theirs except for our limited rate of churn.
They show that they can determine the size of the  network in $O(n^2)$ rounds
and also give a method for approximate counting.
 We differ in that our bounds are sublinear in $n$ (when $D$ is small) and we  maintain our dense graphs at all times.

We work under the well-studied CONGEST model (see, e.g., \cite{peleg} and the references therein). Because of its realistic communication restrictions, there has been much research done in this model (e.g., see \cite{lynch,peleg,PK09}).  In particular, there has been much work done in designing very fast distributed approximation
algorithms (that are even faster at the cost of producing sub-optimal solutions) for many fundamental problems (see, e.g., \cite{elkin-survey,dubhashi,khan-disc,khan-podc}).
Among many graph problems studied, the densest subgraph problem falls into the ``global problem'' category where it seems that one needs at least $\Omega(D)$ rounds to compute or approximate (since one needs to at least know the number of nodes in the graph in order to compute the density). While most results we are aware of in this category were shown to have a lower bound of $\Omega(\sqrt{n/\log n})$, even on graphs with small diameter (see \cite{DasSarmaHKKNPPW11} and references therein), the densest subgraph problem is one example for which this lower bound does not hold.

Our algorithm requires certain size estimation algorithms as a subroutine. An important tool that also addresses network size estimation is a \emph{Controller}.
%
%
Controllers were introduced in~\cite{AfekAPSController-FOCS87}  and they were implemented on `growing' trees, but this was later extended to a more general dynamic model~\cite{KormanKuttenControllerPODC07,EmekKormanController-DISC09}. Network size estimation itself is a fundamental problem in the distributed setting and closely related to other problems like leader election.  For anonymous networks and under some reasonable assumptions, exact size estimation  was shown to be impossible~\cite{CidonS-IPL95} as was leader election~\cite{AngluinSTOC80} (using symmetry concerns). Since then, many probabilistic estimation techniques have been proposed using exponential and geometric distributions~\cite{KuhnLO10,AggarwalKuttenFST93,MatiasA-WDAG89}. Of course, the problem is even more challenging in the dynamic setting.
%

%
%
%
%
Self-* systems~\cite{Berns09DissectingSelf-*,Djikstra74SelfStabilizing,DolevBookSelfStabilization,KormanKMPODC11,Kuhn2005-Repairing,Poor-SelfHealQueue2003,Ghosh07Self-healingSystemsSurvey,PanduranganPODC11,HayesFG-DCJournal-springerlink,HayesPODC09,Amitabh-2010-PhdThesis} are worth mentioning here.
Often, a crucial condition for such systems is the initial detection of a particular state. In this respect, our algorithm can be viewed as a self-aware algorithm where the nodes monitor their state with respect to the environment, and this could be used for developing powerful self-* algorithms.

\section{Future Work and Conclusions}\label{sec:conc}

We have presented efficient decentralized algorithms for finding dense subgraphs in distributed dynamic networks. Our algorithms not only show how to compute size-constrained dense subgraphs with provable approximation guarantees, but also show how these can be {\em maintained} over time. While there has been significant research on several variants of the dense subgraph computation problem in the classical setting, to the best of our knowledge this is the first formal treatment of this problem for a distributed peer-to-peer network model.

Several directions for future research result naturally out of our work. The first specific question is whether our algorithms and analyses can be improved to guarantee $O(D + \log n)$ rounds instead of $O(D\log n)$, even in static networks. Alternatively, can one show a lower bound of $\Omega(D\log n)$ in static networks? Bounding the value $D$ in terms of the instantaneous graphs and change rate $r$ would also be an interesting direction of future work.
It is also interesting to show whether the densest subgraph problem can be solved {\em exactly} in $O(D\poly\log n)$ or not in the static setting, and to develop dynamic algorithms without density lower bound assumptions. Another open problem (suggested to us by David Peleg) that seems to be much harder is the {\em at-most-$k$ densest subgraph problem}.
One could also consider various other definitions of density and study distributed algorithms for them, as well as explore whether any of these techniques extend directly or indirectly to specific applications. Finally, it would be interesting to extend our results from the edge alteration model to allow node alterations as well.

\bibliography{dense,selfheal}

\appendix
\section*{Appendix}
\section{Counting the number of nodes and edges in a subgraph}
\label{sec:count}

Our algorithms make use of an operation in which all the nodes (edges) in a given subgraph need to be counted for different phases of the algorithm. We achieve this by using the node-counting algorithm of Kuhn et al.\cite[Algorithm 2]{KuhnLO10} that gives a $(1 \pm \epsilon)$-approximation of the number of nodes in a network. There are, however, several modifications that have to be made to their algorithm to work in our setting, and we describe these next.

For completeness, our modified version of Kuhn et al.'s algorithm is given in Algorithm~\ref{algo:kuhn}. Note that their algorithm requires an upper bound on the size of the network ($N$), the very quantity that we are estimating. We later give an algorithm that can provide this upper bound, thereby removing this assumption. This algorithm works by generating a number of independent exponential variables at each node and using the fact that the minimum of such quantities gives a means for estimating their cardinality. The first change we make is that we do not have the entire network run this algorithm in a given phase, but only the nodes in the current subgraph (denoted here as $V'$). Though all the nodes in the nework take part in the computation, only the nodes in $V'$ generate exponentially distributed values and hence the final estimate is for this subgraph. Secondly, we change the termination condition of the algorithm. The algorithm of Kuhn et al.\ terminated when a reasonable estimate was reached at each node. Since in our context we have a bound on the number of rounds it takes a message to traverse the network (the dynamic diameter $D$), we simply run for this many rounds and are guaranteed that by the end of $D$ rounds all the nodes have the same minimum values. The proof that this algorithm gives a $(1 \pm \epsilon)$-approximation with high probability is nearly identical to that in \cite{KuhnLO10}, and is hence omitted here.

\begin{algorithm}[h]
\caption{\cite{KuhnLO10}{\sc RandomizedApproximateCounting}($V'$, $D$, $\epsilon$)}
\label{algo:kuhn}
{\bf Input:} A set of nodes $V' \subseteq V$ (each node knows whether it is in $V'$ or not), dynamic diameter $D$, and error parameter $\epsilon$.\\
{\bf Output:} $n'$, a $(1 \pm \epsilon)$-approximation of the number of nodes in $V'$.
\begin{algorithmic}[1]
\STATE Let $c > 0$ and  let $N$ be an upper bound on the size of the network
\STATE Let $l = \lceil 27(2 + 2c)\log{N}/\epsilon^2 \rceil$
\STATE Each node $u \in V'$ generates an $l$-tuple of independent exponential variables with rate 1: $Z^u = (Y^u_1, \ldots, Y^u_l)$; all other nodes $v \in V - V'$ generate $Z^v = (\infty, \infty, \ldots, \infty)$.
\FOR{$r$ = $1, \ldots, D$}
\STATE Broadcast $Z^u$ if $Z^u \neq (\infty, \infty, \ldots, \infty)$.
\STATE Receive $Z^{v_1}, \ldots, Z^{v_s}$ from neighbors.
\FOR{$i = 1, \ldots, l$}
\STATE $Z^u_i = \min{\{Z^u_i, Z^{v_1}_i, \ldots, Z^{v_s}_i\}}$
\ENDFOR
\ENDFOR
\STATE Output $n_u = l/\sum_{i=1}^l Z^u_i$.
\end{algorithmic}
\end{algorithm}

As was noted above, the algorithm of Kuhn et al.\ needs an upper bound $N$ on the size of the network. In Algorithm~\ref{algo:count nodes}, we give an algorithm that provides this upper bound (indeed, a $2$-approximation) using a similar technique. It does not assume that nodes have unique IDs nor does it need an upper bound on the size of $V'$, but it does need to know the dynamic diameter $D$. The algorithm is similar to the ELECT algorithm in~\cite{AfekM94}, except that we use it here to estimate the size of a set of nodes in a dynamic network rather than elect a leader in a static one. This algorithm uses the maximum (rather than the minimum) of discrete (rather than real-valued) independent exponentially distributed values.

\begin{algorithm}[h]
\caption{{\sc Count Nodes}($V'$, $D$, $\delta$)}
\label{algo:count nodes}
{\bf Input:} A set of nodes $V'$ (each node knows whether it is in $V'$ or not), dynamic diameter $D$, and a failure probability $\delta$.\\
{\bf Output:} $n'$, a $(2, \delta)$-approximation of the number of nodes in $V'$ (i.e., if the number of nodes in $V'$ is $n$, then $P(n/2 \leq n' \leq 2n) > 1 - \delta$)
\begin{algorithmic}[1]
\STATE Let $l = 65\ln{(1/\delta)}$
\FOR{$i = 1, \ldots, l$}
\STATE Each node $v \in V'$ tosses an unbiased coin until it sees a head. Let $X^v_i$ be the number of tosses it performs.
\ENDFOR
\FOR{$r = 1, \ldots, D$}
\STATE Broadcast $X^v$ to all of its neighbors.
\STATE Receive $X^{v_1}, \ldots, X^{v_s}$ from neighbors.
\FOR{$i = 1, \ldots, l$}
\STATE $X^v_i = \max{\{X^v_i, X^{v_1}_i, \ldots, X^{v_s}_i\}}$
\ENDFOR
\ENDFOR
\STATE Output the median of $(2^{X^v_1}, \ldots, 2^{X^v_l})$.
\end{algorithmic}
\end{algorithm}

The estimation guarantee of the algorithm is given by the following theorem:
\begin{theorem}[Approximation guarantee] After $D$ rounds, all the nodes have the same estimate of $n = |V'|$ and this estimate $n'$ is such that $P(n/2 \leq n' \leq 2n) > 1 - \delta$.
\end{theorem}
\begin{proof}
Consider any one coordinate of the $l$-tuple, say $i$. After $D$ rounds, by the definition of dynamic diameter, all the values $X^v_i$ have been transmitted to all the nodes, and so they all have the same maximum value. We show that $2^{X^v_i}$ is a good approximation of $n$.

For an arbitrary $X^v_i$, we have a cumulative distribution function of $P(X^v_i \leq k) = (1 - 1/2^k)$. Hence, the cumulative distribution function of $X_{max} = max_{v \in V'} X^v_i$ is $P(X_{max} \leq k) = (1 - 1/2^k)^n$. From this we can compute the probability:
\begin{align*}
P(\lg{n} - 1 \leq X_{max} \leq \lg{n} + 1) &= P\left(X_{max} \leq \lg{n} + 1\right) - P\left(X_{max} \leq \lg{n} - 2\right) \\
&=\left(1 - \frac{1}{2n}\right)^n - \left(1 - \frac{4}{n}\right)^n  \\
&> 1/2 \hfil \qquad \text{(for $n \geq 4$).} 
\end{align*}
Hence, $n/2 \leq 2^{X_{max}} \leq 2n$ with probability greater than $1/2$. Using standard Chernoff bound techniques, it is easy to show that taking the median of $l = O(\ln{(1/\delta)})$ such estimates reduces the failure probability down to $\delta$.
\end{proof}

Note that since all the nodes know the value of $D$, Algorithm~\ref{algo:count nodes} takes precisely $D$ rounds to execute. Also note that the maximum number of bits that a node has to transmit per round is not too high. We can bound $X_{max}$ to within $O(\log{n})$ with high probability, and so no node communications more than $O(\log{(1/\delta)}\log{\log{n}})$ bits in a given round with high probability.

In summary, in each phase of our algorithm we use Algorithm~\ref{algo:count nodes} to get an upper bound on the size of $V'$ (with high probability) and then apply the modified algorithm of Kuhn et al.\ (Algorithm~\ref{algo:kuhn}) to get a $(1 \pm \epsilon)$-approximation of the size of $V'$ using the upper bound from the previous algorithm, all in precisely $2D$ rounds of communication.

We next discuss how the number of edges in the induced subgraph is computed. The algorithm for counting edges is based on the one for counting the number of nodes: each node in the subgraph $u$ counts its degree $d_u$ and simulates the behavior of Algorithms~\ref{algo:count nodes} and \ref{algo:kuhn} with $d_u$ independent copies of the exponentially distributed tuples. This increases the computation cost at each node by a $d_u$ factor, but doesn't affect the number of rounds for the above algorithms. Also note that since the component-wise max or min of the tuples is all that gets transmitted, there is no increase in the amount of data being broadcast by each node. At the end of the computation, the nodes have an estimate of two times the number of edges in the subgraph (since both nodes at the end of an edge report it). The details are given in Algorithm~\ref{algo:count edges}.

\begin{algorithm}
\caption{{\sc Count Edges}($V'$, $\epsilon$)}
\label{algo:count edges}
{\bf Input:} A set of nodes $V'$ (each node knows whether it is in $V'$ or not) and number $\epsilon>0$.\\
{\bf Output:} The algorithm computes $m'$, a $(1 \pm \epsilon)$-approximation to the number of edges in $V'$.
\begin{algorithmic}[1]
\STATE Every node in $V'$ broadcasts a message to its neighbors.
\STATE Each node $u$ counts the number of neighbors in $V'$ that communicated with it, call this $d_u$.
\STATE Algorithm~\ref{algo:count nodes} is run, with each node $u$ simulating $d_u$ separate nodes, to get an upper bound on $\sum_u d_u$.
\STATE Algorithm~\ref{algo:kuhn} is run, with each node $u$ simulating $d_u$ separate nodes, to get a $(1 \pm \epsilon)$ estimate of $\sum_u d_u$, call it $m'$.
\STATE Output $m'/2$
\end{algorithmic}
\end{algorithm}

The analysis of the approximation guarantee for the number of edges is almost identical to that for the number of nodes, and is omitted here. Counting the number of edges also takes $2D$ rounds in total, with no node broadcasting more than $O(\log{n})$ bits in any round with high probability.

\section{Proof of Theorem~\ref{thm:atleastktheorem}}
\label{sec:app main two}

%

\begin{lemma}\label{lem:t prime prime to OPT for at least k}
$\rho_{t''}(V_i\cup \hat{V})>\frac{1-\delta}{3(1+\delta)}\min\left(\frac{\rho_{t'}(H^*)}{1+\delta}, \rho_{t'}(H^*)-3\delta \rho_t(H^*)\right)\,.$
\end{lemma}
\begin{proof}
Let $H'$ be the at-least-$k$ densest subgraph of $G_{t'}$. Note that
\begin{align}\label{eq:at least k one}
\rho_{t'}(H^*)\leq \rho_{t'}(H')\,.
\end{align}

Now, define $\ell$, $H^1, \ldots, H^\ell$ and $D$ using Algorithm~\ref{algo:algo for proof} (which is similar to the process defined in \cite{KS} to prove that the algorithm in \cite{KS} is a 2-approximation). We note that we are not interested in the efficiency of this algorithm as it is only used to prove the approximation guarantee.
\begin{algorithm}
\caption{Defining $\ell$, $H^1, \ldots, H^\ell$ and $D$ for the proof of Lemma~\ref{lem:t prime prime to OPT for at least k}.} \label{algo:algo for proof}
\begin{algorithmic}[1]
\STATE Let $j=0$, $G_{t'}^0=G_{t'}$ and $D=\emptyset$. For any set of vertices $X$, let $E_{t'}(X)$ be the set of edges in the subgraph of $G_{t'}$ induced by $X$.
\WHILE{$|D|<k/(1-\delta)$ or $|E_{t'}(D)\cap E_{t'}(H')|< \frac{1}{3}E_{t'}(H')$}
\STATE For any $j$, let $H^j$ be the densest subgraph of $G_{t'}^j$.
\STATE $D=D\cup V(H^j)$.
\STATE Let $G_{t'}^{j+1}$ be the graph obtained from $G_{t'}^j$ by deleting nodes in $H^j$.
\STATE $j=j+1$.
\ENDWHILE
\STATE Let $\ell=j-1$.
\end{algorithmic}
\end{algorithm}

Note the following simple observation:
\begin{observation}\label{obs:48}\label{observation:proof of at least k}
For all $j=1, \ldots, \ell$, $\rho_{t'}(H^j)\geq \frac{2}{3}\rho_{t'}(H').$
\end{observation}
\begin{proof}
Since $|E_{t'}(D)\cap E_{t'}(H')|< \frac{1}{3}E_{t'}(H')$  in every iteration of the while loop,
$$|E_{t'}\left(V(G_{t'}^j)\cap V(H')\right)|\geq \frac{2}{3}|E_{t'}(H')|\,.$$
That is, there are at least $2/3$ fraction of edges of $H'$ left in $G_{t'}^j$. This implies that the density of subgraph of $G_{t'}^j$ induced by nodes in $H'$ is at least
\begin{eqnarray*}
 \rho_{t'}\left(V(G_{t'}^j)\cap V(H')\right) = \frac{|E_{t'}\left(V(G_{t'}^j)\cap V(H')\right)|}{|V(G_{t'}^j)\cap V(H')|} 
\geq  \frac{2}{3}\frac{|E_{t'}(H')|}{|V(H')|}  
= \frac{2}{3}\rho_{t'}(H')\,. 
\end{eqnarray*}
Since $H^j$ is the densest subgraph of $G_{t'}^j$,

$$
\rho_{t'}(H^j)\geq \rho_{t'}\left(V(G_{t'}^j)\cap V(H')\right)  \geq  \frac{2}{3}\rho_{t'}(H')$$
as claimed.
\end{proof}


Let $i^*$ be the smallest index such that $V(D)\subseteq V_{i^*}$ and $V(D)\not\subseteq V_{i^*+1}$. Note that $i^*$ exists since the algorithm repeats until we get $V_j=\emptyset$. Now we consider two cases.



{\bf Case 1: $n_{i^*}\geq k$.} Let $v$ be any vertex in $V(D)\setminus V_{i^*+1}$. Let $j^*$ be such that $v\in V(H^{j^*})$. Note that Observation~\ref{observation:proof of at least k} implies that
\begin{align}\label{eq:at least k one one}
\rho_{t'}(H')\leq \frac{3}{2}\rho_{t'}(H^{j^*})\,.
\end{align}
Let $H_{t', i^*}$ be the subgraph of $G_{t'}$ induced by vertices in $V_{i^*}$. Note that
\begin{align}\label{eq:at least k one two}
\rho_{t'}(H^{j^*})\leq 2\deg_{H^{j^*}}(v) \leq 2\deg_{H_{t', i^*}}(v)\,.
\end{align}
%
The first inequality is because we can remove $v$ from $H^{j^*}$ and get a subgraph of $G_{t'}^{j^*}$ that has higher density than $H^{j^*}$ otherwise. The second inequality is because $H^{j^*}\subseteq H_{t', i^*}$ (since $V(H^{j^*})\subseteq D\subseteq V_{i^*}$).
%
%
%
%
Since $v$ is removed from $V_{i^*}$,
\begin{align}\label{eq:at least k one three}
\deg_{H_{t',i^*}}(v)< (1+\delta) \frac{m_{i^*}}{n_{i^*}}
\end{align}
where $\delta=\epsilon/24$ as in Algorithm~\ref{algo:maintain}. By definition of $i$ and the fact that $n_{i^*}\geq k$,
\begin{align}\label{eq:at least k one four}
\frac{m_{i^*}}{n_{i^*}}=\frac{m_{i^*}}{\max(k, n_{i^*})}\leq \frac{m_i}{\max(k, n_i)}\,.
\end{align}
%
Note that $|V_i\cup \hat{V}|\leq n_i/(1-\delta)$ and $m_{i}\leq (1+\delta) |E_{t''}(V_{i})|$ with high probability\danupon{Need proof from previous section.}. It follows that
\begin{align}\label{eq:at least k one five}
\frac{m_i}{\max(k, n_i)}\leq \frac{1+\delta}{1-\delta}\rho_{t''}(V_{i}\cup \hat{V})\,.
\end{align}
Combining Eq.\eqref{eq:at least k one} with \eqref{eq:at least k one one}-\eqref{eq:at least k one five}, we get $\rho_{t'}(H^*)<3\frac{(1+\delta)^2}{1-\delta}\rho_{t''}(V_i\cup \hat{V})$\danupon{To polish: carefully take care of $\hat{V}$} and thus the lemma.



{\bf Case 2: $n_{i^*}<k$.} This implies that with high probability $|V_{i^*}|<(1+\delta)k$. Since $D\subseteq V_{i^*}$, $|D|<(1+\delta)k$. By the condition in the while loop of Algorithm~\ref{algo:algo for proof},
\begin{align}\label{eq:at least k two one}
|E_{t'}(V_{i^*})| \geq |E_{t'}(D)| \geq \frac{1}{3}|E_{t'}(H')|\,.
\end{align}
%
%
Note that $m_{i^*}\geq |E_{t'}(V_{i^*})|-Tr\geq \frac{1}{3}|E_{t'}(H')|-\delta k \rho_t(H^*)$.
%
Thus,
\begin{align}\label{eq:at least k two two}
\frac{m_{i^*}}{\max(k, n_{i^*})} &\geq \frac{\frac{1}{3}|E_{t'}(H')|-\delta k \rho_{t}(H^*)}{k} 
\geq \frac{1}{3}\rho_{t'}(H')-\delta \rho_t(H^*)\,.
\end{align}
By Eq.\eqref{eq:at least k one},
\begin{align}\label{eq:at least k two three}
\frac{m_{i}}{\max(k, n_{i})} &\geq \frac{m_{i^*}}{\max(k, n_{i^*})} 
\geq \frac{1}{3}\rho_{t'}(H')+\delta \rho_t(H^*) 
\geq \frac{1}{3}\rho_{t'}(H^*)-\delta \rho_t(H^*)\,.
\end{align}
%
%
Note that $|V_i\cup \hat{V}|\leq k/(1-\delta)$ and $m_{i}\leq (1+\delta) |E_{t''}(V_{i})|$ with high probability\danupon{Need proof from previous section.}. It follows that
\begin{eqnarray*}\label{eq:at least k two five}
\frac{m_i}{\max(k, n_i)} & \leq  \frac{(1+\delta)|E_{t''}(V_{i})|}{(1-\delta) k} \leq  \frac{1+\delta}{1-\delta}\rho_{t''}(V_{i}\cup \hat{V})\,.
\end{eqnarray*}
%
Combining Eq.\eqref{eq:at least k one}, \eqref{eq:at least k two three} and \eqref{eq:at least k two five}, we get  $\rho_{t''}(V_i\cup \hat{V})>\frac{(1-\delta)(\rho_{t'}(H^*)-3\delta \rho_t(H^*))}{3(1+\delta)}$ and thus the lemma.
\end{proof}

\subsection{Proof of Observation~\ref{obs:48}}
\label{sec:app3}

\begin{proof}
Since $|E_{t'}(D)\cap E_{t'}(H')|< \frac{1}{3}E_{t'}(H')$  in every iteration of the while loop,
$$|E_{t'}\left(V(G_{t'}^j)\cap V(H')\right)|\geq \frac{2}{3}|E_{t'}(H')|\,.$$
That is, there are at least $2/3$ fraction of edges of $H'$ left in $G_{t'}^j$. This implies that the density of subgraph of $G_{t'}^j$ induced by nodes in $H'$ is at least
\begin{eqnarray*}
\rho_{t'}\left(V(G_{t'}^j)\cap V(H')\right) & = & \frac{|E_{t'}\left(V(G_{t'}^j)\cap V(H')\right)|}{|V(G_{t'}^j)\cap V(H')|}\\
 & \geq & \frac{2}{3}\frac{|E_{t'}(H')|}{|V(H')|}\\
&  = & \frac{2}{3}\rho_{t'}(H')\,.
\end{eqnarray*}
Since $H^j$ is the densest subgraph of $G_{t'}^j$,
$$\rho_{t'}(H^j)\geq \rho_{t'}\left(V(G_{t'}^j)\cap V(H')\right)\geq \frac{2}{3}\rho_{t'}(H')$$
as claimed.
\end{proof}

\begin{proof} [Proof of Theorem~\ref{thm:atleastktheorem}]
The theorem follows directly by using Lemma~\ref{lem:t prime prime to OPT for at least k} and translating the lower bound on $\rho_{t''}(V_i\cup \hat{V})$ to a lower bound on $\rho_{t}(V_i\cup \hat{V})$ (similar to the steps in proof of Theorem~\ref{theorem:approx densest}). As can be seen, the Lemma~\ref{lem:t prime prime to OPT for at least k} has a factor $\frac{1-\delta}{3(1+\delta)}$ as compared to a similar term of $\frac{(1-\delta)^2}{2(1+\delta)^2}$ in the case for densest subgraph. This is why we are only able to obtain a $(3+\epsilon)$-approximation to this theorem rather than a $(2+\epsilon)$-approximation previously. The proof for the theorem and the $(3+\epsilon)$-approximation is completed as before by translating $\rho_{t''}(V_i\cup \hat{V})$ to a lower bound on $\rho_{t}(V_i\cup \hat{V})$ and subsequently from the $\frac{1-\delta}{3(1+\delta)}$ term by plugging in the appropriate value for $\delta$ in terms of $\epsilon$.
\end{proof}

\end{document}